\documentclass[12pt]{article}	
\def\Eq#1{Eq.\ (\ref{#1})}
\def\Eqs#1#2{Eqs.\ (\ref{#1}) and (\ref{#2})}
\def\vec#1{\mbox{\boldmath{$#1$}}}
\def\maxeq4{Eqs.\ (\ref{divE}--\ref{curlB})}

\title{\bf The road not considered ... the question of photon mass}

\author{\bf Palash B. Pal\\ 
\normalsize Saha Institute of Nuclear Physics,
1/AF Bidhan-Nagar, Calcutta 700064, INDIA}

\date{November 2005}

\begin{document}

\maketitle

It is well-known that a conflict between Galilean relativity and
Maxwell equations of electromagnetic equations was apparent by the end
of the 19th century.  Equally well-known is the fact that Einstein
resolved the conflict by pointing out that the principle of relativity
and Maxwell equations can both be correct if the notion of absolute
time is abandoned.  The rest is history.  The purpose of this article
is not to recount that history.  Rather, we wonder about a question
which, it seems, should have been asked in the process, but apparently
was not.

Newtonian formulation of dynamics was firmly established by the
nineteenth century, proving its worth in all sorts of questions that
were addressed within its framework.  Einstein's introduction of
Special Theory of Relativity required a reformulation of the laws of
dynamics.  Newtonian dynamics became only an approximation, valid in
the regime of small speeds.  Maxwell equation of electrodynamics,
however, remained unaffected.

And this is where the question comes.  In the times of great
upheavals, it is natural to re-examine everything conventional.
Newtonian dynamics was indeed re-examined and replaced by a more
sophiticated alternative with the advent of special relativity.  Why
wasn't there any serious discussion about any modification or
reformulation of Maxwell equations at the same time?

It almost seems like a vain question.  Wasn't the entire conflict
created by the presence of the quantity $c$ in Maxwell equations,
implying that the speed of electromagnetic waves in the vacuum is
independent of the observer?  And didn't Einstein absorb this
statement as a postulate in his system, thus assuming the immutability
of Maxwell equations in the basic structure of his Special Theory?

Yes, it is undeniable that the Maxwell equations contained a
fundamental quantity which has the dimension of velocity.  This is
most apparent if we write the equations in Gaussian units, where the
electric and the magnetic fields are assumed to have the same
dimension.  In this system of units, the equations are:
\begin{eqnarray}
\vec\nabla \cdot \vec E &=& 0 \,, 
\label{divE}\\*
\vec\nabla \cdot \vec B &=& 0 \,,  
\label{divB}\\
\vec\nabla \times \vec E + {1\over c} 
{\partial \vec B \over \partial t} &=& 0 \,, 
\label{curlE}\\* 
\vec\nabla \times \vec B - {1\over c} {\partial \vec E \over \partial
t} &=& 0  \,.
\label{curlB}
\end{eqnarray}
Note that we have written the equations in absence of any source,
which is all that we will need in this article.  The constant $c$,
carrying the dimension of velocity, explicitly appears even in the
absence of sources.

In other systems of units, this constant may be hidden.  For example,
if we write down the Maxwell equation involving the sources in SI
units, the equations apparently contain two ``fundamental'' constants
$\varepsilon_0$ and $\mu_0$, called respectively the permittivity and
permeability of free space, and none of them have the dimension of
velocity.  However, there is nothing fundamental about the ratio
$\varepsilon_0/\mu_0$: it merely serves to set the relative units
between the charge density and the current density.  If no source is
present, only the combination $\varepsilon_0\mu_0$ appears in the
Maxwell equations, and it is easily seen that this combination has the
dimension of inverse velocity squared.  In fact, it equals the
quantity $1/c^2$, where $c$ appears in \Eqs{curlE}{curlB} written in
the Gaussian system of units.

It therefore follows that this quantity $c$ is a fundamental constant,
and it should therefore be observer-independent, and Einstein boldly
recognized this fact by extending the Galilean principle of relativity
to it.  But should this $c$ be the speed of the electromagnetic waves?

The answer is `yes' if one accepts \maxeq4 without any modification
anywhere, as we all learn from textbooks on electromagnetic theory.
This is trivial to see.  We merely have to take the curls of the last
two equations, applying the vector calculus identity
\begin{eqnarray}
\vec\nabla \times (\vec\nabla \times \vec V) = \vec\nabla (\vec\nabla
\cdot \vec V) - \vec\nabla^2 \vec V 
\label{curlcurl}
\end{eqnarray}
which holds for any differentiable vector field $\vec V$.  After that,
using \Eqs{divE}{divB}, we readily obtain
\begin{eqnarray}
{1\over c^2} {\partial^2 \vec E \over \partial
t^2} - \nabla^2 \vec E &=& 0 \,, 
\label{waveE}\\*
{1\over c^2} {\partial^2 \vec B \over \partial
t^2} - \nabla^2 \vec B &=& 0 \,,
\label{waveB}
\end{eqnarray}
which are wave equations, with the speed of the waves equal to $c$.
Since $c$ is a universal constant, \Eqs{waveE}{waveB} imply that the
speed of electromagnetic waves is independent of its wavelength.  In
other words, the vacuum must be a dispersion-free background for the
propagation of electromagnetic waves.

This is where the question comes.  Didn't that look awkward to anyone
in the beginning of the twentieth century?  The phenomenon of
dispersion was known since Newton's famous experiment with prism, and
in the long experience since then, no one has ever encountered a
medium that is dispersion-free.  So, why didn't the idea of a
dispersion-free vacuum appear outrageous?  Instead, wouldn't it have
been more natural to suppose that the quantity $c$ that appears in the
Maxwell equations is invariant, but it is not the speed of the
electromagnetic waves, and that the Maxwell equations somehow have to
be modified so that \Eqs{waveE}{waveB} do not follow from them?

The question becomes more poignant in the light of the fact that in
1905, Einstein also worked with the hypothesis of light quanta or
photons.  From his pioneering paper on the special theory of
relativity, he already knew that for a particle of mass\footnote{In
older texts on relativity, this quantity was called the ``rest mass'',
in order to distinguish it from the ratio $E/c^2$ of a moving
particle.  I will follow the modern use of the term ``mass'', which
signifies a Lorentz invariant property, equal to $E_0/c^2$, where
$E_0$ is the energy of the particle at rest.}  $m$, the relation
between energy and momentum is given by
\begin{eqnarray}
E^2 = \vec p^2 c^2 + m^2 c^4 \,.
\end{eqnarray}
Using Hamilton's dynamical equations, one can then easily deduce that
for a given momentum, the velocity of the particle is given by
\begin{eqnarray}
\vec v = {\partial E \over \partial \vec p} = {\vec p c \over
  \sqrt{\vec p^2 + m^2 c^2}} \,.
\end{eqnarray}
This shows two things about the magnitude of velocity.  First, it
depends on the magnitude of momentum, which means that there is
dispersion.  Second, the magnitude of velocity is less than $c$ for
any non-zero value of the mass.  In order to agree with the
dispersionless speed of electromagnetic waves obtained from
\Eqs{waveE}{waveB}, one therefore has to conclude that $m=0$, i.e., the
photons are massless.

A century has passed since the publication of Einstein's papers on
special relativity and light quanta.  In these hundred years, we have
grown so accustomed to this statement about the masslessness of the
photon that we seldom realize that it must have been a mysterious
statement in those days, and is no less mysterious now.  All known
material particles have some mass.  The speed of any particle depends
on its momentum.  Any particle can be at rest with respect to a given
observer.  If we want to think of photons as particles, it is natural
to assign the same set of properties to them.  And yet, this
supposedly ``natural'' road was never considered.  Why not?  We
discuss possible answers to this question in the rest of this article.

Certainly the phenomenological success of the electromagnetic theory
was not the reason behind the supposed immutability of Maxwell's
equations.  After all, Newtonian mechanics also had tremendous
phenomenological success, and was in fact tested for a much longer
time in history.  So, if Newtonian mechanics could have been modified,
Maxwellian electrodynamics could have been challenged as well.

One might be tempted to think that the Maxwell equations embodied the
beautiful structure of gauge invariance, and no one would have tried
to modify the equations and spoil the gauge invariance.  This seems to
be a very unlikely explanation.  In 1905, the idea of the photon was
just coming into light (no pun intended).  Quantum electrodynamics
(QED) was still a few decades away into the future.  There was no way
to guess that gauge invariance would play an important role in QED.
As far as classical physics was concerned, gauge invariance was only a
mathematical curiosity, which was useful only in solving the
differential equations of electrodynamics with less trouble.  There
was no big physical significance attached to the idea of gauge
invariance, and therefore no physicist treasured the idea as something
very fundamental.

A third guess would be the following.  Einstein had to use the idea of
the constancy of the speed of light, because he used it to develop the
entire structure of special relativity, starting from the relativity
of simultaneity.  This is the basic reason why he, and others of his
time, did not want to disturb the equation for electromagnetic wave
given in \Eqs{waveE}{waveB}.  The sentiment is understandable, but
this was not the only logical possibility.  As we emphasized earlier,
Maxwell's equations necessarily contained one constant which has the
dimension of velocity.  It was a bold decision to acknowledge that
this fundamental quantity must be observer-independent, and this was
the big step taken by Einstein.  Once this step is taken, it is
trivial to argue from there that the concept of time must be
observer-dependent, and this was the second step that Einstein took.
But after that, one can just use the principle of relativity, and some
basic axioms like that of the homogeneity of space and time, to arrive
at \cite{nothingbut} the Lorentz transformation equations and the
relativistic velocity addition rule.  One needs to use the fact that
$c$ is a constant, but nowhere one has to identify it with the speed
of light in the vacuum.

Let us then explore how Maxwell's equation could have been modified.
We will assume that the modified equation would still be linear in the
electric and magnetic fields, and we will keep considering the
sourceless equations.  One possibility is to keep the divergence
equations, \Eqs{divE}{divB}, unchanged, and modify the curl equations:
\begin{eqnarray}
\vec\nabla \times \vec E &=& - {1\over c} 
{\partial \vec B \over \partial t} + \eta_1 \vec E + \xi_1 \vec B\,, 
\label{curlE0}\\* 
\vec\nabla \times \vec B &=& {1\over c} {\partial \vec E \over \partial
t} + \eta_2 \vec E + \xi_2 \vec B \,,
\label{curlB0}
\end{eqnarray}
where $\eta_1,\eta_2$, $\xi_1,\xi_2$ are constants which cannot be
independent, as will be discussed soon.  Taking the curls of these two
equations and using \Eq{curlcurl}, we obtain
\begin{eqnarray}
- \vec\nabla^2 \vec E &=& - {1\over c} {\partial \over \partial t}
  \vec\nabla \times \vec B + \eta_1 \vec\nabla \times \vec E + \xi_1
  \vec\nabla \times \vec B \,.
\end{eqnarray}
Using \Eqs{curlE0}{curlB0} for the curls of the electric and magnetic
fields on the right hand side and rearranging the terms, we get
\begin{eqnarray}
{1\over c^2} {\partial^2 \vec E \over \partial t^2} - \vec\nabla^2 \vec E 
&=& {\xi_1 - \eta_2\over c} {\partial \vec E \over \partial t}
   - {\xi_2 + \eta_1 \over c}  {\partial \vec B \over \partial t} 
\nonumber\\* && 
+ (\eta_1^2 + \xi_1 \eta_2) \vec E +
  \xi_1 (\eta_1 + \xi_2) \vec B \,. 
\end{eqnarray}
This is a mess, and a similar mess occurs if we try to find the
equation containing second derivatives of the magnetic field.  First,
the equations are coupled in electric and magnetic fields.  Second,
they contain first order time derivative terms, which would produce
damped solutions which should not occur in absence of any dissipation.
Fortunately, curing the second problem is easily done by choosing
\begin{eqnarray}
\xi_1 = \eta_2 \,, \qquad \xi_2 = -\eta_1 \,.
\end{eqnarray}
And then we note that it automatically cures the first problem.
However, even after this, the equation turns out to be
\begin{eqnarray}
{1\over c^2} {\partial^2 \vec E \over \partial t^2} - \vec\nabla^2
\vec E = (\eta_1^2 + \eta_2^2) \vec E \,.
\label{badwaveeq}
\end{eqnarray}
Since the electric and magnetic fields are real, the constants
$\eta_1,\eta_2$ must also be real.   In this case, \Eq{badwaveeq}
gives unphysical solutions.  For example, if a field is homogeneous,
\Eq{badwaveeq} says that its magnitude must be exponentially
increasing or decreasing in the vacuum.  This does not make any sense.

So let us try something different.  Again, we will modify two of the
four equations of \maxeq4, but this time we choose a different
grouping.  If we introduce the sources, it would affect only two of
the four equations.  The other two, viz., \Eqs{divB}{curlE}, are just
constraint equations that the electric and magnetic fields must
satisfy.  Let us keep these constraints equations undisturbed, but
modify the other two by writing
\begin{eqnarray}
\vec\nabla \cdot \vec E &=& f \,, 
\label{divEf}\\*
\vec\nabla \times \vec B &=& {1\over c} {\partial \vec E \over \partial
t} + \vec F  \,,
\label{curlBF}
\end{eqnarray}
where obviously $f$ is a scalar and $\vec F$ is a vector.  Note that
these are not the sources.  We have written the equation in the
vacuum, where there are no sources, and we assumed that even then
there are some other terms in the equation.  We will soon comment on
the nature of these terms.

We can take the curl of both sides of \Eq{curlBF}, use the identity of
\Eq{curlcurl} and then \Eq{divB} to write
\begin{eqnarray}
{1\over c^2} {\partial^2 \vec B \over \partial t^2} - \vec\nabla^2
  \vec B = \vec\nabla \times \vec F \,.
\end{eqnarray}
What needs to be done if we want to obtain a meaningful wave equation
for the magnetic field?  Obviously, we want the right hand side to be
proportional to $\vec B$.  The proportionality constant must have the
dimension of inverse length squared.  If this constant is positive, we
get into the kind of problems discussed in connection with
\Eq{badwaveeq}.  Therefore, we want
\begin{eqnarray}
\vec\nabla \times \vec F = - {1 \over \ell^2} \vec B \,,
\label{curlF}
\end{eqnarray}
where $\ell$ is a constant carrying the dimension of length, and the
equation for the magnetic field in the vacuum becomes
\begin{eqnarray}
{1\over c^2} {\partial^2 \vec B \over \partial t^2} - \vec\nabla^2
  \vec B = - {1 \over \ell^2} \vec B \,.
\end{eqnarray}
The same equation would be satisfied by the electric field as well,
provided we identify
\begin{eqnarray}
- \vec\nabla f - {1\over c}{\partial \vec F \over \partial t} 
= - {1 \over \ell^2} \vec E \,.
\label{gradf}
\end{eqnarray}

Purely within the purview of classical physics, this would have opened
a new Pandora's box.  It would seem that now we are in the need of a
fundamental quantity with the dimension of length.  But since Einstein
already possessed the idea of a light quantum, that was not really
necessary.  In fact, with the help of Planck's constant, one could
have written
\begin{eqnarray}
\ell = {\hbar \over mc} \,,
\end{eqnarray}
where $m$ is something with the dimension of mass.  And this mass need
not have been any fundamental constant: it could have been interpreted
as the mass of the photon.  

So, to come back to the question: why wasn't this modification of
Maxwellian electrodynamics put to test?  Although there is no way to
be sure about the answer, one can make a strong guess.  Remember that
the homogeneous Maxwell equations have been kept unchanged in this
formulation, so we can define the potentials through the usual
relations:
\begin{eqnarray}
\vec B &=& \vec\nabla \times \vec A \,, \nonumber\\*
\vec E &=& - \vec\nabla\varphi - {1\over c}{\partial \vec A \over
  \partial t} \,. 
\end{eqnarray}
Now, looking back at \Eqs{curlF}{gradf}, we realize that 
\begin{eqnarray}
\vec F &=& - (mc/\hbar)^2 \vec A \,, \nonumber\\*
f &=& - (mc/\hbar)^2 \varphi \,.
\end{eqnarray}
Thus the modified Maxwell equations would contain the scalar and the
vector potentials.

This is something that would have seemed like a complete catastrophe
for a physicist in 1905.  As we already said, the potentials were
supposed to be only mathematical constructs which could help solve the
dynamical equations.  They were not supposed to enter any equation of
physical significance.  And maybe that is why modification of Maxwell
equations in the line described above was never considered.

This is not to say that no modification of Maxwellian electrodynamics
was at all considered at the time when special relativity was
introduced.  Indeed, there was the emission theory of Ritz \cite{ritz}
and others.  But they violated Lorentz transformation equations.  And
it is understandable that, having obtained the Lorentz transformation
equations and having them justified by Einstein's effort, it was
physchologically not easy to discard them altogether.  We raise the
question about a milder alternative: the Lorentz transformation
equations are not disturbed, but the quantity $c$ that appears in them
is not the speed of light in the vacuum.

The reader must have noticed that we have always stuck to the
principles and the machinery available to a physicist of 1905, and not
any hindsight that has developed since that time.  For example, the
modification attempted in \Eqs{curlE0}{curlB0} can easily be discarded
today, by noting that the modified equations cannot be expressed in a
covariant 4-dimensional notation.  On the other hand, the modification
introduced through \Eqs{divEf}{curlBF} do have a simple representation
in 4-dimensional notation, and are indeed the field equations derived
from the Proca Lagrangian \cite{proca}.  In modern day language, the
question raised in this article can be summarized in this form: why
was the Proca equation introduced as late as in 1930?  Why didn't one
start from the Proca equations right in 1905 and allowed the photon
mass to be determined by experiments?  Of course, even at the outset
anyone would have known that the photon mass must be very small, and
further experiments could have been designed to know the mass better.
Indeed, a very similar path has been pursued for neutrinos.  Though at
first it was proposed that neutrinos are massless, soon after the
possibility of its mass was taken seriously, experiments were
performed to determine the mass, and through decades of effort we now
know that neutrino mass, albeit very small compared to all other known
fermion masses, is indeed non-zero.  This program, which seemed very
logical and pragmatic, was not undertaken for the photon at the
beginning of the twentieth century, when a lot of thought went into
the nature of light and electromagnetic theory.  Of course, data
collected for other reasons were used later to put bounds on photon
mass \cite{massrev}.

There is an interesting irony in these set of events.  Quantum theory
developed quite independently of special relativity.  We can fantasize
over things that could have happened if, instead of beginning their
journeys almost at the same time in the hands of Einstein, quantum
theory had started much before the theory of relativity.  What if
relativity flourished after the Aharanov-Bohm theory and the
experiments to confirm it, which showed that electromagnetic
potential, even in the absence of electromagnetic field, can affect
outcomes of physical experiments?  Probably in such circumstances, no
one would consider the presence of potentials in equations of motion
outrageous, and the question of photon mass would have been raised
with the advent of relativity theory.

On hindsight, however, it seems good that things happened the way they
did.  With the advent of quantum theory, it was recognized that gauge
invariance might serve as a crucial ingredient in constructing the
quantum theory of electromagnetic interactions or QED.  Even later, it
was recognized that the same principle, suitably generalized, holds
the key to the construction of gauge theories, which now describe the
theories of strong and weak interactions as well.  It would have been
a pity if gauge invariance, present in Maxwell's equations, was
sacrificed by introducing the possibility of a photon mass.

I thank P. DasGupta for discussions and suggestions for improvement.


\begin{thebibliography}{[10]}

\bibitem{nothingbut} See, e.g.,\\ N. D. Mermin: {\sl Relativity
without light}, Am.  J. Phys. 52 (1984) 119; \\ P. B. Pal: {\sl
Nothing but Relativity}, Eur. J. Phys. 24 (2003) 315.

\bibitem{ritz} W. Ritz: Ann. Chim. Phys. 13 (1908) 145.  For reviews
  and references, see, e.g., J. G. Fox: Rev. Mod. Phys. 33 (1965) 1.

\bibitem{proca} A. Proca: Compt. Rend. 190 (1930) 1377; 191 (1930) 26.

\bibitem{massrev} For a review and references, see, e.g.,
  A. S. Goldhaber and M. M. Nieto: Rev. Mod. Phys. 43 (1971) 277.

\end{thebibliography}
\end{document}